\documentstyle[prl,aps,epsfig]{revtex}
\begin{document}
\draft \title{Square vortex lattices for two component superconducting 
order parameters}
\author{D.F. Agterberg}
\address{Theoretische Physik, Eidgen\"{o}ssiche Technische Hochschule-
H\"{o}nggerberg, 8093 Z\"{u}rich, Switzerland}
\date{\today}
\maketitle
\begin{abstract}
I investigate the vortex lattice structure of the 
Ginzburg Landau free energy for a two component 
order parameter in the weak-coupling clean-limit when the field
is along the high symmetry axis in a tetragonal crystal. 
It is shown that the vortex lattice phase diagram as a function
of the Ginzburg Landau free energy parameters includes phases
with a hexagonal, centered rectangular, rectangular, and square
unit cells. It is also shown
that the square vortex lattice has the largest region of stability.
The field 
distribution of the square vortex lattice near $H_{c2}$ is determined and the
application of this model to Sr$_2$RuO$_4$ is discussed. 
\end{abstract}

The oxide Sr$_2$RuO$_4$ has a structure similar to high $T_c$ materials
and was discovered to be superconducting with a $T_c=1.35$ K by 
Maeno {\it et. al} in 1994 \cite{mae94}. 
It has been established that this superconductor is not a conventional
$s$-wave superconductor:  
NQR measurements
show no indication of a Hebel-Slichter peak in $1/T_1T$ \cite{ish97},
$T_c$ is strongly suppressed by non-magnetic impurities \cite{mac98},
and tunneling experiments are inconsistent with $s$-wave
pairing \cite{jin98}. More recently there have been two experimental 
results that shed more light on the nature of the superconducting state.
The $\mu$SR experiments of Luke {\it et. al.} 
indicate that the superconducting state {\it breaks} time reversal 
symmetry which 
implies that the superconducting 
order parameter must have more than one component \cite{luk98}.
Of the possible representations (REPS) of the $D_{4h}$ point group, 
the two dimensional (2D) $\Gamma_{5u}$ representation (REP) 
is the most likely 
state that exhibits this property. 
The order parameter in this case
has two components $({\eta_1,\eta_2})$
that share the same rotation-inversion  symmetry
properties as $(k_x,k_y)$ \cite{sig91}. The broken ${\cal T}$ state would
then correspond to $(\eta_1,\eta_2)\propto(1,i)$.
Theoretical arguments supporting a triplet pairing state have been 
given in Ref.~\cite{ric95}.
Given that this material may well be described by such an order parameter,
it is of interest to explore further consequences of a two-component 
order parameter.  In an earlier work  it was shown that a consequence
of the low temperature broken time reversal symmetry state 
is that the mean field vortex lattice phase diagram will  
exhibit two vortex lattice phases when the field is along a high 
symmetry direction in the basal plane \cite{agt98}. 
Another important experimental development is the observation of
a square vortex lattice in Sr$_2$RuO$_4$ by Forgan {\it et. al}
\cite{for98}. Within the context of the orbital dependent superconductivity
model for Sr$_2$RuO$_4$ the orientation of the square vortex lattice 
relative to the underlying ionic lattice dictates which of the Ru orbitals
exhibit superconductivity \cite{agt98}.   
This work focuses
on the magnetic field distribution and the structure
of the vortex lattice for the field
along the $c$-axis for this
two-component model.

The free energy for the $\Gamma_{5u}$ representation of the tetragonal 
point group is given by \cite{sig91} 
\begin{eqnarray}
f=&-\alpha|\vec{\eta}|^2+\beta_1|\vec{\eta}|^4/2+
\beta_2(\eta_1\eta_2^*-\eta_2\eta_1^*)^2/2
+\beta_3|\eta_1|^2|\eta_2|^2 +
\kappa_1(|\tilde{D}_x\eta_1|^2+|\tilde{D}_y\eta_2|^2)\label{eq1}\\
&+\kappa_2(|\tilde{D}_y\eta_1|^2+
|\tilde{D}_x\eta_2|^2)+ \kappa_5(|\tilde{D}_z\eta_1|^2+|\tilde{D}_z\eta_2|^2)
\nonumber \\ & +\kappa_3[(\tilde{D}_x\eta_1)(\tilde{D}_y\eta_2)^*+h.c.]+
\kappa_4[(\tilde{D}_y\eta_1)(\tilde{D}_x\eta_2)^*+h.c.]
+h^2/(8\pi).
\nonumber
\end{eqnarray}
where $\alpha=\alpha_0(T-T_c)$, $\tilde{D}_j=\nabla_j-\frac{2ie}{\hbar c} A_j$,
${\bf h}=\nabla\times {\bf A}$, and ${\bf A}$ is the vector potential. 
To simplify the analysis the Ginzburg Landau coefficients are
determined within a weak-coupling approximation in the clean limit.
The measurements
of Mackenzie {\it et. al.} of $T_c$ as a function of impurity concentration
show that the ratio of the mean free path to the zero-temperature
coherence length is $>8$ for $T_c>1.3$ K \cite{mac98} indicating that the clean
limit should be a reasonable approximation for Sr$_2$RuO$_4$.
Without an experimental knowledge of the characteristic frequency of 
the boson responsible for the pairing
(presumably ferromagnetic spin fluctuations) it cannot be determined 
that the weak-coupling limit is appropriate for Sr$_2$RuO$_4$. 
Note that the spin fluctuation theory  of Mazin and Singh indicate 
that $T_c/T_P\approx 10^{-2}$ where $T_P$ is the characteristic 
paramagnon frequency \cite{maz97}. This estimate in conjunction with  
$T_c/T_F\approx 10^{-4}$ indicates that the weak-coupling approximation 
is  reasonable for Sr$_2$RuO$_4$, but further
experiments are required to ensure this.
Taking for the $\Gamma_{5u}$ REP the gap function
described by the pseudo-spin-pairing gap matrix (note that this choice is not
unique):
$\hat{\Delta}=i[\eta_1 v_x/(\langle v_x^2\rangle)^{1/2}+
\eta_2v_y/(\langle v_x^2\rangle)^{1/2}]\sigma_z \sigma_y$
where the brackets $\langle \rangle$ denote an average over the Fermi 
surface
and $\sigma_i$ are the Pauli matrices, 
writing $\eta_{+}=(\eta_1+ i\eta_2)/\sqrt{2}$, 
$\eta_-=(\eta_1-i\eta_2)/\sqrt{2}$, and rotating 
$(\tilde{D}_x,\tilde{D}_y)$ by an angle $\theta$ about the $z$-axis
to obtain $(\tilde{D}_{\tilde{x}},\tilde{D}_{\tilde{y}})$, 
the following dimensionless free energy is found 
\begin{eqnarray} 
f=&-(|\eta_+|^2+|\eta_-|^2)+(|\eta_+|^4+|\eta_-|^4)/2
+2|\eta_+|^2|\eta_-|^2+\nu[(\eta_-\eta_+^*)^2/2
+(\eta_-^*\eta_+)^2]\label{eq2}\\&
+|D_{\tilde{x}}\eta_+|^2+|D_{\tilde{y}}\eta_-|^2 +|D_{\tilde{x}}\eta_-|^2+
|D_{\tilde{y}}\eta_+|^2 \nonumber\\
&+(e^{i2\theta}+\nu
e^{-i2\theta})[(D_{\tilde{x}}\eta_+)(D_{\tilde{x}}\eta_-)^*
-(D_{\tilde{y}}\eta_+)(D_{\tilde{y}}\eta_-)^*]/2 
\nonumber \\&+(e^{-i2\theta}+\nu
e^{i2\theta)})[(D_{\tilde{x}}\eta_-)(D_{\tilde{x}}\eta_+)^*
-(D_{\tilde{y}}\eta_-)(D_{\tilde{y}}\eta_+)^*]/2
\nonumber \\
&+ I(e^{-i2\theta}-\nu
e^{i2\theta})[(D_{\tilde{x}}\eta_-)(D_{\tilde{y}}\eta_+)^*
+(D_{\tilde{y}}\eta_-)(D_{\tilde{x}}\eta_+)^*]/2
\nonumber \\& -I(e^{i2\theta}-\nu
e^{-i2\theta})[(D_{\tilde{x}}\eta_+)(D_{\tilde{y}}\eta_-)^*
+(D_{\tilde{y}}\eta_+)(D_{\tilde{x}}\eta_-)^*]/2 
+\tilde{\kappa}_5(|D_z\eta_+|^2+|D_z\eta_-|^2)+h^2 \nonumber,
\end{eqnarray}
where $h=\nabla\times {\bf A}$, 
$D_{\nu}=\nabla_{\nu}/\kappa-iA_{\nu}$, 
$f$ is in units $B_c^2/(4\pi)$, lengths are 
in units $\lambda=[\hbar^2 c^2 \beta_1/(32 e^2 
\kappa_1\alpha \pi)]^{1/2}$,
$\tilde{\kappa}_5=2\kappa_5/\kappa_{12}$,
$\kappa_{12}=\kappa_1+\kappa_2=4\kappa_1/(3+\nu)$,
$\nu=(\langle v_x^4 \rangle -3\langle v_x^2v_y^2 \rangle )/
(\langle v_x^4 \rangle+ \langle v_x^2v_y^2\rangle )$,
$h$ is in units $\sqrt{2}B_c=\Phi_0/(2\pi\lambda\zeta)$
(here $B_c$ has been chosen to represent the thermodynamic
critical field),
$\alpha=\alpha_0(T-T_c)$, $\xi=(\kappa_{12}/2\alpha)^{1/2}$, 
and $\kappa=\lambda/\xi$. The parameter $\nu$ (note $|\nu|\le 1$) gives a 
measure of the square anisotropy of the Fermi surface. For
a cylindrical Fermi surface $\nu=0$ and for a square Fermi surface
$|\nu|=1$.
It is easy to verify that in zero-field  
$(\eta_1,\eta_2)\propto(1,i)$  is the stable ground
state for $|\nu|\le 1$.

\begin{figure}
\epsfxsize=100mm
\centerline{\epsffile{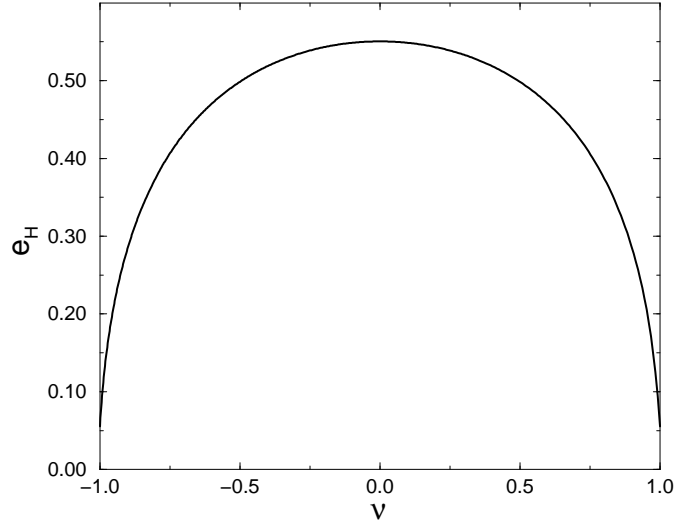}}
\caption{$e_H=\kappa/H_{c_2}$ as a function of
$\nu$.}
\label{fig1}
\end{figure}

For the magnetic field along the $c$-axis the ground state is found by setting
$D_z=0$. Writing $\Pi_+=\sqrt{\kappa}(iD_{\tilde{x}}+D_{\tilde{y}})/\sqrt{2H}$ and
$\Pi_-=\sqrt{\kappa}(iD_{\tilde{x}}-D_{\tilde{y}})/\sqrt{2H}$,  
minimizing the quadratic
portion of Eq.~\ref{eq2} with respect to $\eta_+$ and $\eta_-$ 
yields
\begin{equation}
\kappa \pmatrix{\eta_+\cr \eta_-}=H\pmatrix{1+2N
&e^{-2i\theta}\Pi_+^2+\nu e^{2i\theta}\Pi_-^2\cr 
e^{2i\theta}\Pi_-^2+\nu e^{-2i\theta}\Pi_+^2&
1+2N}\pmatrix{\eta_+\cr \eta_-}\label{eq5}.
\end{equation}
where $N=\Pi_+\Pi_-$. The maximum value of the externally applied field 
$H$ that allows a non-zero
solution for $(\eta_+,\eta_-)$ yields the upper critical field $H_{c_2}$.
At $H=H_{c2}$ the vector potential is that for a spatially uniform field
${\bf A}=(0,Hx,0)$. 
Expanding
$(\eta_+,\eta_-)$ in terms of the eigenstates of $N$ (Landau levels)
up to $N=32$
and diagonalizing the
resulting matrix yields the result for $e_H=\kappa/H_{c_2}$ shown
as a function of $\nu$ 
in Fig.~\ref{fig1}.

\begin{figure}
\epsfxsize=70mm
\centerline{\epsffile{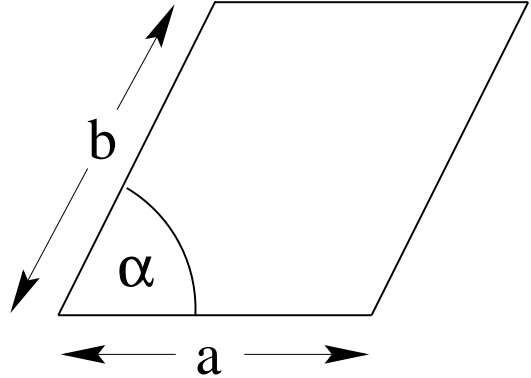}}
\caption{The vortex lattice unit cell}
\label{fig2}
\end{figure}

The form of the eigenstate
of the $H_{c_2}$ solution is found to be
$\eta_+({\bf r})=\sum_{n\ge 0} a_{4n+2} \phi_{4n+2}({\bf r})$ and
$\eta_-({\bf r})=\sum_{n\ge 0} a_{4n} \phi_{4n}({\bf r})$ where
$\phi_n({\bf r})$ are the harmonic oscillator wave functions (Landau levels). 
As is well known these wave functions have a large degeneracy, and the
form of the vortex lattice is found by including the nonlinear terms 
of the Ginzburg Landau equations perturbatively to break this degeneracy.
Following the procedure of Abrikosov, the average Gibbs free energy density 
is found to be
(a derivation of this result for unconventional superconductors can 
be found in Ref.~\cite{luk95} and for a mixed $d$ and $s$-wave order parameter
in Ref.~\cite{fra96}) 
\begin{equation}
\overline{g}=-H^2-\frac{(H_{c_2}-H)^2}{(2\tilde{\kappa}^2-1)\beta_A}
\end{equation}
where
\begin{equation}
\beta_A=
\frac{\overline{h_s^2}}{(\overline{h_s})^2},
\end{equation}
\begin{equation}
2\tilde{\kappa}^2=\frac{\overline{f_4}}{\overline{h_s^2}}
\end{equation}
$f_4$ is the fourth order homogeneous free energy, $h_s$ is the 
field (along the $c$-axis) induced by the supercurrent, and the over-bar
denotes a spatial average. 
The form of the vortex lattice
is found by minimizing $(2\tilde{\kappa}^2-1)\beta_A$. 
To do this $h_s$ must be found.
By minimizing the Ginzburg Landau free energy with respect to the
vector potential the following relation is found for $h_s$
\begin{eqnarray}
j_+\equiv \frac{\partial h_s}{\partial y}-i\frac{\partial h_s}{\partial x}=&
\eta_+^*(\Pi_-\eta_+)+\eta_+(\Pi_+\eta_+)^*+\eta_-^*(\Pi_-\eta_-)
+\eta_-(\Pi_+\eta_-)^*
\label{eq6}\\ &
+e^{i2\theta}[\eta_-^*(\Pi_+\eta_+)+\eta_+(\Pi_-\eta_-)^*]
+\nu e^{i2\theta}[\eta_+^*(\Pi_+\eta_-)+\eta_-(\Pi_-\eta_+)^*]
\nonumber 
\cr
j_-\equiv \frac{\partial h_s}{\partial y}+i\frac{\partial h_s}{\partial x}=
&
\eta_+(\Pi_-\eta_+)^*+\eta_+^*(\Pi_+\eta_+)+\eta_-(\Pi_-\eta_-)^*
+\eta_-^*(\Pi_+\eta_-)\nonumber \cr &
+e^{-i2\theta}[\eta_-(\Pi_+\eta_+)^*+\eta_+^*(\Pi_-\eta_-)]
+\nu e^{-i2\theta}[\eta_+(\Pi_+\eta_-)^*+\eta_-^*(\Pi_-\eta_+)]
\nonumber 
\end{eqnarray}
Near $H_{c_2}$ the field $h_s$ is found by substituting the vector potential 
for the homogeneous
field and the order parameter solution near $H_{c_2}$ 
into the right hand side of Eq.~\ref{eq6}. 
Writing the left hand side of Eq.~\ref{eq6} as  
$j_i=\sum_{n,m} (j_i)_{n,m}
\phi_n\phi_m^*$  and writing  
$h_s=\sum_{n,m}h_{n,m}\phi_n\phi_m^*$ yields (see Ref~\cite{tak71})
\begin{eqnarray}
h_{l,l}
=&\sum_{n=l}^{\infty}(j_+)_{n+1,n}/\sqrt{n+1}& \label{hs} \\
h_{l,p}=&[\sqrt{l}(j_+)_{l-1,p}-\sqrt{p}(j_+)_{p-1,l}^*]/(p-l)
&\hphantom{abcdefghijk}l\ne p.
\nonumber
\end{eqnarray}
The form of $\beta_A$ and $\tilde{\kappa}$ will 
therefore be determined by terms of the type 
$\overline{\phi_n({\bf r})\phi_m^*({\bf r})\phi_p({\bf r})\phi_l^*({\bf r})}$.
To evaluate such terms I make the assumption that the vortex lattice unit
cell contains one flux quantum. The shape of the
unit cell is kept arbitrary and the convention introduced by Saint-James
{\it et. al} \cite{sai69} 
to describe the unit cell is used. The lattice geometry is depicted in 
Fig.~\ref{fig2}.

The lattice vectors are ${\bf a}_1=a(1,0)$ and ${\bf a}_2=b(\cos \alpha,
\sin \alpha)$ with the single flux quantum constraint 
$ab\sin\alpha=2\pi$ where $a$ and $b$ are in units
$l_H=2\pi\sqrt{\lambda \xi/H}$.
This assumption allows the functions $\phi_n$ to be written as
\cite{luk95}
\begin{equation}
\phi_n({\bf r})=2^{-n/2}\pi^{-1/4}(n!)^{-1/2}\sum_{m}c_me^{i2\pi(m-1/2)x/a}e^{-(y-y_m)^2/2}H_n(y-y_m)
\end{equation} 
where $c_n=e^{i\pi n(\rho+1-n\rho)}$, $y_n=(n-1/2)\sqrt{2\pi\sigma}$,
$\sigma=(b/a)\sin\alpha$, $\rho=(b/a)\cos\alpha$,
and $H_n$ are the Hermite polynomials.
To evaluate $\beta_A$ and $\tilde{\kappa}$ 
it is useful to express the spatial 
averages in terms of a sum over the reciprocal lattice of
the vortex lattice. The reciprocal lattice is given by 
${\bf G}=l_1{\bf k}_1+l_2{\bf k}_2$ where
 ${\bf k}_1=\frac{2\pi}{a \sin\alpha}(\sin\alpha,-\cos\alpha)$
${\bf k}_2=\frac{2\pi}{b \sin\alpha}(1,0)$.  
The general form of $\beta_A$ becomes 
\begin{equation}
 \beta_A=\frac
{\sum_{n,m,p,l} a_{n,m,p,l}\sum_{l_1,l_2}\langle \phi_n \phi_m^*\rangle_{l_1,l_2}\langle \phi_p
\phi_l^*\rangle_{-l_1,-l_2}}{[\sum_n a_n \langle |\phi_n|^2 \rangle_{0,0}]^2}
\end{equation} where the coefficients $a_{n,m,p,l}$ and $a_n$ are determined
by $f_4$, $h_s$, and the form of the eigenfunction near $H_{c_2}^c$,
 $\langle f \rangle_{l_1,l_2}=(ab\sin\alpha)^{-1}\int_{uc} e^{-i{\bf G}
\cdot {\bf r}} f({\bf r})$, and $uc$ denotes the unit cell.
The  following relation makes this formulation for determining $\beta_A$
and $\tilde{\kappa}$ useful (found using the addition theorem for Hermite
polynomials)
\begin{equation}
\langle \phi_p \phi_m^* \rangle_{l_1,l_2}=
\frac{2^{-(m+p)/2}}{\sqrt{p!m!}}
\langle |\phi_0|^2 \rangle_{l_1,l_2}\sum_{r=0}^p\sum_{s=0}^m\frac{C_{r,s}
H_{p-r}[-z_{l_1,l_2}]H_{m-s}[z_{l_1,l_2}^*]}{(p-r)!(m-s)!}
\end{equation} 
where 
\begin{equation}
C_{r,s}=\sum_{l=0}^{[r/2]}\sum_{p=0}^{[s/2]}(-1)^{l+p}\frac{
(r+s-2l-2p)!2^{-l-p+(r+s)/2}}{(r-2l)!(s-2p)!l!p![(r+s)/2-l-p]!},
\end{equation}  
$z_{l_1,l_2}=\sqrt{\pi}[l_1 \sqrt{\sigma}+i(l_2-\rho l_1)/\sqrt{\sigma}]$,
and
\begin{equation}
\langle |\phi_0|^2 \rangle_{
  l_1,l_2}=\frac{\sqrt{\pi}}{b\sin\alpha}e^{i\pi(l_1+l_2+l_1l_2)}
e^{-\pi \sigma l_1^2/2}e^{-\pi(l_2-l_1\rho)^2/2\sigma}. 
\end{equation}
The following relation is also useful
\begin{equation}
\langle \phi_p \phi_0^* \rangle_{l_1,l_2}= [\sqrt{2}z_{l_1,l_2}]^p
\langle |\phi_0|^2 \rangle_{l_1,l_2}.
\end{equation}
The relations 11-14 are straightforward
to implement numerically.

In the analysis of the form of the vortex lattice the parameter $\nu$ 
was considered to lowest order in perturbation theory (recall that $|\nu|\le1$
so that $\nu$ provides a natural expansion parameter).
The limit $\nu=0$ was considered by Zhitomirsky who analytically found
the ground state eigenvector near $H_{c_2}$ \cite{zhi89}.  
The solution of the order
parameter to first order in $\nu$ is
\begin{equation}
(\eta_+,\eta_-)=[\phi_0+b_4\nu e^{-i4\theta}\phi_4,-e^{i2\theta}(\epsilon
\phi_2+b_6\nu e^{-i4\theta}\phi_6)]
\end{equation}
where $\epsilon=\sqrt{3}-\sqrt{2}\approx 0.31784$, 
$b_4=2\sqrt{3}\epsilon(10+\sqrt{6})/(36+16\sqrt{6})\approx
0.18230$, and $b_6=\sqrt{30}b_4/(10+\sqrt{6})\approx 0.080203$.
 Substituting this solution for the eigenstate
 near $H_{c_2}$ into the Eq. 6 yields the coefficients
\begin{eqnarray}
(j_+)_{0,1}=&1-\sqrt{2}\epsilon\cr
(j_+)_{0,5}=&e^{i4 \theta}\nu(\sqrt{5}b_4-\sqrt{6}b_6) \nonumber\cr
(j_+)_{1,2}=&\sqrt{2}\epsilon^2-\epsilon\nonumber\cr
(j_+)_{1,6}=&e^{i4 \theta}b_6\nu (\sqrt{2}\epsilon-1)\nonumber\cr
(j_+)_{2,3}=&\sqrt{3}\epsilon^2\nonumber\cr
(j_+)_{2,7}=&e^{i 4 \theta}\sqrt{7}\epsilon b_6 \nu\nonumber\cr
(j_+)_{3,0}=&e^{i4 \theta}\nu(2b_4-\sqrt{3}\epsilon)\nonumber\cr
(j_+)_{4,1}=&e^{-i4 \theta}\nu b_4(1-\sqrt{2}\epsilon)\nonumber\cr
(j_+)_{5,2}=&e^{-i4 \theta}\epsilon\nu(\sqrt{6}b_6-\sqrt{5}b_4)\nonumber\cr
(j_+)_{6,3}=&e^{-i4 \theta}b_6\nu\sqrt{3}\epsilon\nonumber
\end{eqnarray}
Application of Eq.~\ref{hs} yields 
\begin{eqnarray}
h_s=&\frac{-1}{2\kappa}[(1-3/\sqrt{2}\epsilon+2\epsilon^2)|\phi_0|^2
+(2\epsilon^2-\epsilon/\sqrt{2})|\phi_1|^2+\epsilon^2|\phi_2|^2 \label{hs2}\\
&+\nu[(\sqrt{10}\epsilon b_4/4-\sqrt{6}b_6/4)(e^{i4\theta}\phi_1^*\phi_5+
e^{-i4\theta}\phi_5^*\phi_1)\cr &
+(\sqrt{30}\epsilon b_4/4-\epsilon b_6
-\sqrt{2}b_6/4)(e^{i4\theta}\phi_2^*\phi_6+e^{-i4\theta}\phi_6^*\phi_2)
\cr & +(\sqrt{3}\epsilon/2-b_4)(e^{i4\theta}\phi_4^*\phi_0
+e^{-i4\theta}\phi_0^*\phi_4)].
\end{eqnarray}
Using this expression for $h_s$, $(2\tilde{\kappa}^2-1)\beta_A$ 
should be minimized with respect
to $\theta, \sigma,$ and $\rho$
to find the form of the vortex lattice.
It can be proven when $\nu> 0$ ($\nu <0$) $(2\tilde{\kappa}^2-1)\beta_A$ 
can be minimized for 
$\theta=\pi/4$ ($\theta=0$). For $\nu=0$, $(2\tilde{\kappa}^2-1)\beta_A$ 
is independent of $\theta$. This is
to be expected since $\nu=0$ corresponds to a cylindrically symmetric
Fermi surface. It is also found that $\tilde{\kappa}$ varies 
weakly ($\approx 0.01$) for the different vortex lattice structures
studied in this article (such behavior is also present for 
mixed $d$ and $s$-wave order parameters \cite{fra96}).
While this variation is small it determines the form of the vortex lattice
in the region of $\tilde{\kappa}\approx 1/\sqrt{2}$ and  at small
$\kappa$ the vortex lattice phase diagram becomes quite rich.

The form of the vortex lattice found in the large 
$\kappa$ limit agrees with that found under more
restrictive assumptions in Ref.~\cite{agt98}. 
In this limit the lattice structure depends upon $\nu$. 
The behavior of the vortex lattice as a function of $\nu$ is similar to 
the behavior as a function of temperature
found for borocarbide \cite{dew97} and $d$-wave \cite{ich97,fra96}
superconductors.
For $\nu=0$ a hexagonal lattice is found. As $|\nu|$ increases the lattice 
deforms continuously until $|\nu|=0.0114$. For $0<|\nu|<0.0114$ the vortex
lattice is a centered rectangular lattice as shown in Fig.3
(which can be described by $\theta=\pi/4-\alpha$ and 
$\rho=\cos\alpha$ and $\sigma=\sin\alpha$ where $\pi/3<\alpha<
\pi/2$, $\alpha=\pi/3$
corresponds to the hexagonal lattice and $\alpha=\pi/2$ to the square lattice).
For $|\nu|\ge0.0114$ the vortex lattice is square.
If $\nu\ge 0.0114$ the vortex lattice is
rotated $\pi/4$ with respect to the underlying crystal lattice 
while for $\nu\le -0.0114$ the vortex lattice is aligned with
the underlying crystal lattice. 

\begin{figure}
\epsfxsize=150mm
\centerline{\epsffile{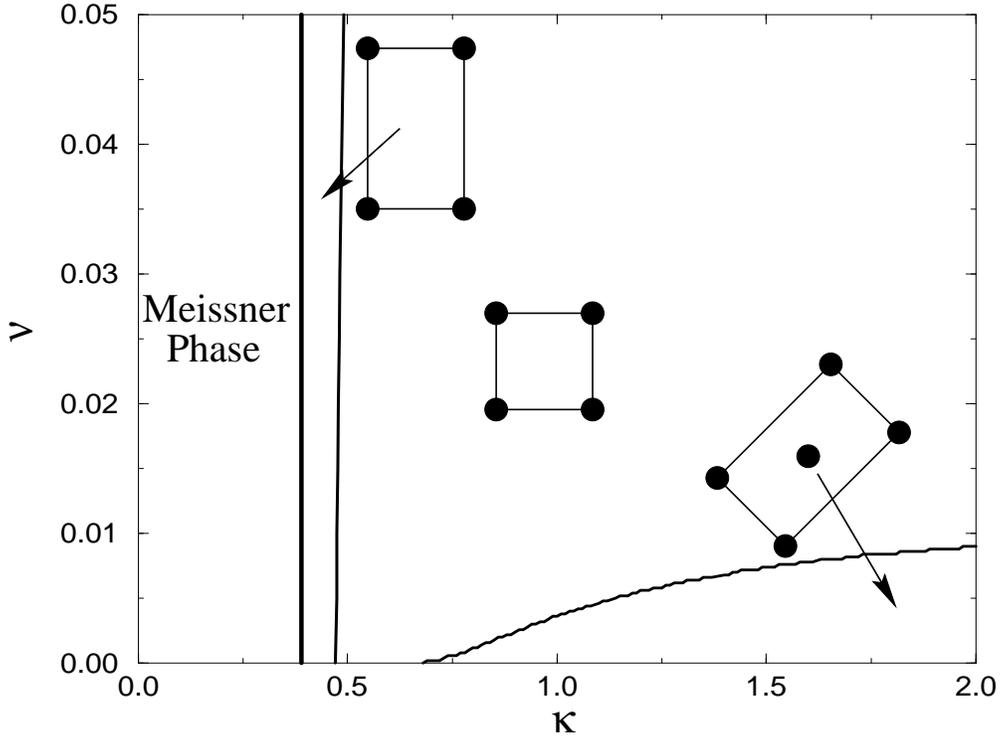}}
\caption{The vortex lattice phase diagram as a function of the
Ginzburg-Landau ratio $\kappa$ and the square anisotropy parameter $\nu$.
The phase diagram is the same for $\nu<0$.
For $\nu>0$ ($\nu<0$)  the square vortex lattice is rotated
$\pi/4$ (0) with respect to the underlying crystal lattice.}
\label{fig3}
\end{figure}

As mentioned above when $\kappa$ becomes sufficiently small
the vortex lattice phase diagram
becomes richer. Fig.~\ref{fig3} shows the region of stability
for the three vortex lattice states that were found to be stable.
In addition to the two phases described above a third phase appears
for small $\kappa$. The vortex lattice for this phase has a
rectangular unit cell and is described by $\rho=0$ and $\sigma=b/a$.
This phase is stable because $\tilde{\kappa}$ is smaller for this
phase than for both the square and hexagonal lattices. 
For a cylindrically symmetric 
Fermi surface ($\nu=0$)
and $\kappa<0.75$ the hexagonal vortex lattice is no longer the stable
structure. This arises because $\tilde{\kappa}$ is in a local
maximum for the hexagonal lattice.

The Type I to Type II transition can also be determined and
it is not given by $\tilde{\kappa}=1/\sqrt{2}$ but by $H_{c_2}=H_c$
(which corresponds to $\kappa= \epsilon\sqrt{3/2}$ up to 
corrections that are second order in $\nu$). For a conventional
superconductor $\tilde{\kappa}=1/\sqrt{2}$ and $H_{c_2}=H_c$ 
are equivalent. Here it is found that the
$\kappa$ for which $\tilde{\kappa}=1/\sqrt{2}$ is less than
$\kappa= \epsilon\sqrt{3/2}$ for all lattice structures studied. 
An analysis of the Gibbs energies indicates
that the Meissner state is the stable phase for $H$ near $H_{c_2}$
when $\kappa<\epsilon\sqrt{3/2}$.

\begin{figure}
\epsfxsize=55mm
\centerline{\epsffile{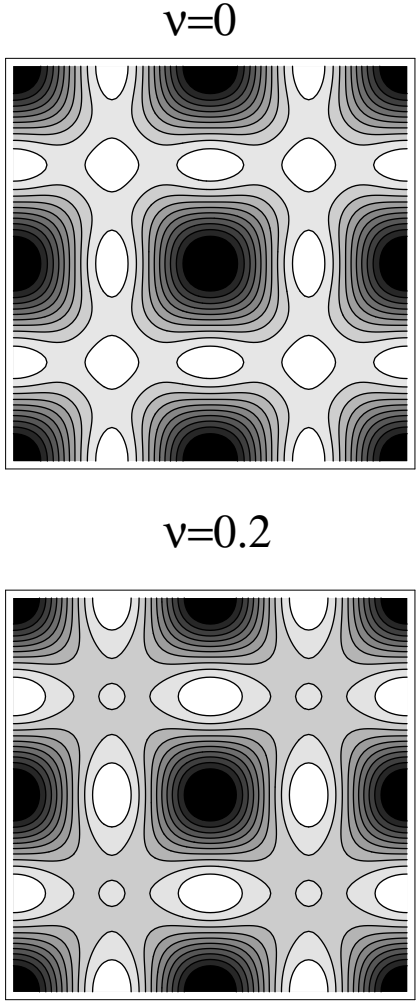}}
\caption{Contour plots of the induced magnetic field $h_s$ for
a square vortex lattice with $\nu=0$ (top) and $\nu=0.2$ (bottom).
The contours from darkest to lightest correspond to $\kappa h_s=-0.42
\rightarrow-0.042$ in
units $0.042$.}
\label{fig4}
\end{figure}

\begin{figure}
\epsfxsize=120mm
\centerline{\epsffile{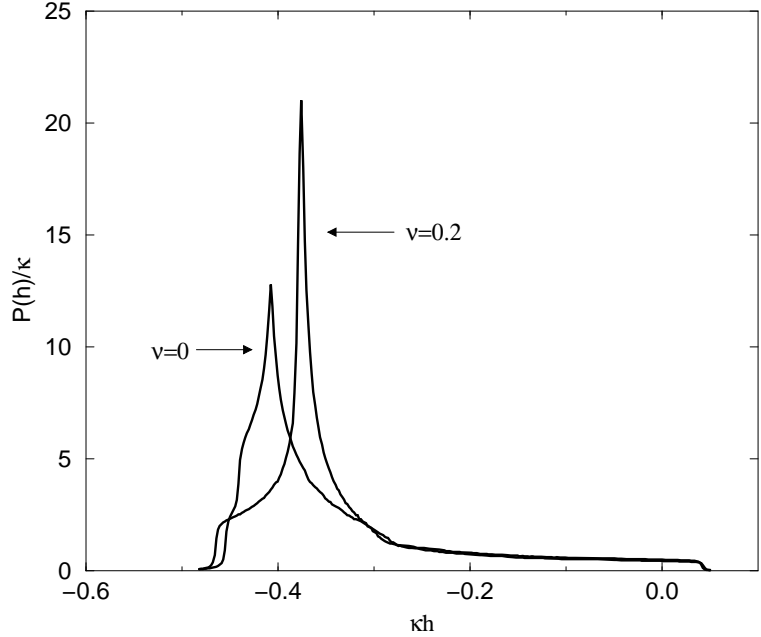}}
\caption{Field distribution $P(h)$ near $H_{c_2}$ for
a square vortex lattice with $\nu=0$ and $\nu=0.2$}
\label{fig5}
\end{figure}

Clearly the square vortex lattice has the largest region of stability
in Fig~\ref{fig3}.
To further investigate the square vortex lattice the spatial variation
of the magnetic field as given by Eq.~\ref{hs2} is determined. This is 
shown in Fig.~\ref{fig4} for $\nu=0$ and for $\nu=0.2$. The induced
field $h_s$ has (in addition to a global maximum and a global minimum)
a local minimum and a saddle point.
Fig.~\ref{fig5} shows the
field distribution  for these two values of $\nu$ as determined 
from
\begin{equation} 
P(h)=\frac{\int d^2r \delta[h-h({\bf r})]}{\int d^2r}.
\end{equation}
The peak in $P(h)$ is due to the saddle point in the spatial
dependence of $h_s$. As $\nu$ increases the saddle point value of $h_s$ 
moves away from the minimum value of $h_s$ resulting in a larger  
'shoulder' in $P(h)$ as $\nu$ increases. 

Now I turn to an application of these results to Sr$_2$RuO$_4$.
This requires a determination of the parameters $\nu$ and $\kappa$.
The value of $\kappa$ as defined above is given by 
\begin{equation}
\kappa=\frac{e_{H} H_{c_2}}{\sqrt{2}H_c},
\end{equation}
 where $e_H$ is given in Fig.~\ref{fig1} and $H_{c}$ ($H_{c_2}$) correspond
to the measured 
critical (upper critical) field [note that the above choice of 
$\kappa$ and $\xi$ 
also implies $H_{c_2}=\Phi_0/(2e_H\pi \xi^2)$].
In principle the values for $H_{c_2}$ and $H_c$ given in Ref.~\cite{yos96}
can be used to estimate $\kappa$, however the sample 
used had a $T_c=0.9$ K which
indicates that impurities cannot be neglected (since $T_{c}^{max}\approx1.5$ K)
so that the clean-limit approximation used here is not valid. 
Until measurements on cleaner samples become available these 
measurements will be used to estimate $\kappa$. 
Using the values of $H_{c_2}$
and $H_c$ given in Ref.~\cite{yos96} yields $\kappa=1.2 e_H<0.7$.
This implies that either the square or the orthorhombic vortex 
lattices will occur depending on the value of $\nu$.

To determine the value of $\nu$ experimentally   
the anisotropy of the upper critical field in the basal
plane can be used \cite{bur85,agt98}
\begin{equation}
\frac{H_{c_2}({\bf a})}{H_{c_2}({\bf a}+{\bf b})}=\frac{1-\nu}{1+\nu}.
\end{equation}
where $H_{c_2}({\bf a})$ [$H_{c_2}({\bf a}+{\bf b})$] is the upper 
critical field for the field along ${\bf a}$ [${\bf a}+{\bf b}$]
measured at $T_c$. This anisotropy has not yet been determined
so that a microscopic model must be used to estimate $\nu$.
 LDA band structure calculations
\cite{ogu95,sin95} reveal that the
density of states
near the Fermi surface is due mainly to the four Ru $4d$ electrons
in the $t_{2g}$ orbitals.
There is a strong hybridization of these orbitals with the O
$2p$ orbitals giving rise to antibonding $\pi^*$ bands. The resulting
bands have three quasi-2D Fermi surface sheets labeled $\alpha,\beta,$
and $\gamma$ (see Ref. \cite{mac296}). The $\alpha$ and $\beta$ sheets
consist of $\{xz,yz\}$ Wannier functions and the
$\gamma$ sheet of $xy$ Wannier functions. 
In general $\nu$
is not given by a simple average over all the sheets
of the Fermi surface.  A knowledge of the pair scattering
amplitude on each sheet and between the sheets is required
to determine $\nu$ \cite{agt97,maz97}.
It has been shown that the pair scattering amplitude
between the $\gamma$ and the $\{\alpha,\beta\}$ sheets of the Fermi 
surface is expected to be small relative to the intra-sheet pair 
scattering amplitudes 
if Sr$_2$RuO$_4$ is not an isotropic $s$-wave superconductor \cite{agt97}.
This forms the basis for the model of orbital dependent superconductivity
in which, to account for 
the large residual density of states observed in the superconducting
state, it has been proposed that either the
$xy$ or the $\{xz,yz\}$ Wannier functions exhibit
superconducting order. This model implies that
that there are two possible values of $\nu$; one for the $\gamma$
sheet ($\nu_{xy}$) and one for an average over the $\{\alpha,\beta\}$
sheets ($\nu_{xz,yz}$). Using the following tight binding dispersions
\begin{eqnarray}
\epsilon_{\gamma}=&\epsilon_{\gamma}^0-2t_{\gamma}(\cos k_x+\cos k_y)
-4\tilde{t}_{\gamma}\cos k_x \cos k_y\\
\epsilon_{\alpha,\beta}=&\epsilon_{\alpha,\beta}^0-2t_{\alpha,\beta}
(\cos k_x +\cos k_y)\pm\sqrt{4t_{\alpha,\beta}^2(\cos k_x -\cos k_y)^2
+16 \tilde{t}_{\alpha,\beta}^2\sin^2 k_x\sin^2k_y}
\nonumber
\end{eqnarray}
and using the tight binding values of Ref.~\cite{maz97} for the
$\gamma$ sheet  
$(\epsilon_{\gamma}^0,t_{\gamma},\tilde{t}_{\gamma})=(-0.4,0.4,0.12)$
and the values 
$(\epsilon_{\alpha,\beta}^0,t_{\alpha,\beta},\tilde{t}_{\alpha,\beta})=
(-0.3,0.25,0.075)$  for the $\{\alpha,\beta\}$ sheets 
yields $\nu_{xy}=-0.6$ and $\nu_{xz,yz}=0.6$.
These values of $\nu$ seem too large since they imply an anisotropy 
of a factor of
4 in $H_{c_2}({\bf a})/H_{c_2}({\bf a}+{\bf b})$.
However $\nu$ depends strongly upon the tight binding parameters used,
for example taking 
$(\epsilon_{\gamma}^0,t_{\gamma},\tilde{t}_{\gamma})=
(-0.52,0.4,0.16)$ yields
$\nu_{xy}=-0.08$. The qualitative result that $\nu_{xy}<0$ and
$\nu_{xz,yz}>0$ is robust. Physically $\nu_{xy}<0$
because of the proximity of the $\gamma$ Fermi surface sheet
to a Van Hove singularity and $\nu_{xz,yz}>0$ due to the quasi
1D nature of the $\{\alpha,\beta\}$ surfaces \cite{ogu95,sin95}.
Assuming $|\nu|<0.2$ implies $\kappa\approx 0.7$ in which case there will
be a square vortex  lattice that is rotated $\pi/4$ (0) with respect to the 
underlying crystal lattice  if the pairing occurs on the $\{\alpha,
\beta\}$ ($\gamma$) Fermi sheets. It is encouraging that Forgan {\it et. al}
have observed a square vortex lattice in Sr$_2$RuO$_4$ \cite{for98}.  
Further experimental studies of the vortex lattice should 
provide useful information as to the nature of the superconducting phase.

In conclusion a Ginzburg Landau  theory
for a two component order parameter representation of the tetragonal 
point group has been examined with a magnetic field
applied along the $c$-axis. The vortex lattice phase diagram near $H_{c_2}$ 
was found to be rich with a square vortex lattice occupying most of the
parameter space. The field distribution of the square vortex lattice 
was determined yielding predictions for $\mu$SR measurements. Finally,
the application of this model to Sr$_2$RuO$_4$ indicates that a square
vortex lattice is expected to appear. The orientation of the
square vortex lattice with respect to the underlying crystal lattice
yields information as to which of the Ru 4$d$ orbitals
are relevant to the superconducting state.

I acknowledge support from the Natural Sciences
and Engineering Research Council of Canada and  the Zentrum for 
Theoretische Physik. I wish to thank E.M. Forgan, R. Heeb, 
G. Luke, A. Mackenzie, 
Y. Maeno, T.M. Rice, and M. Sigrist for useful discussions.

\end{document}